\documentclass[acmsmall, nonacm]{acmart}

 \AtBeginDocument{%
  }

\usepackage{hyperref}
\usepackage{listings}
\usepackage{float}
\usepackage{csquotes}
\usepackage{xcolor}
\usepackage{adjustbox}
\usepackage{microtype}
\usepackage{graphicx}
\usepackage{subcaption}
\usepackage{booktabs} 
\usepackage{amsmath}
\usepackage{mathtools}
\usepackage{amsthm}
\usepackage{stmaryrd}

\newcommand{\nop}[1]{}
\newcommand{\quartz}{\textsc{Covan}}
\newcommand{\qatc}{\textsc{Coat}}

\usepackage{algorithm}
\usepackage{algorithmic}

\usepackage{tikz}
\usepackage{tikz-cd}
\usetikzlibrary{matrix}
\usetikzlibrary{arrows.meta, positioning, bending}

\usepackage{tcolorbox}

\usepackage{natbib}

\setcopyright{none}
\renewcommand\footnotetextcopyrightpermission[1]{} 

\title[Coalgebraic Provenance Tracking]{Provenance Tracking in AI Compilers through the Lens of Coalgebra}
\author{Zilu Tian}
\affiliation{
    \institution{OmniVision Technology}
    \city{Singapore}
    \country{Singapore}
}
\email{zilutian1@gmail.com}
\orcid{0000-0003-4339-1876}

\author{Liying Liu}
\affiliation{
    \institution{Black Sesame Technology}
    \city{Singapore}
    \country{Singapore}
}
\email{liyingliu2010@gmail.com}
\orcid{0009-0000-8072-8461}

\DeclareMathOperator{\Ob}{Ob}
\DeclareMathOperator{\Hom}{Hom}

\usepackage[capitalize,noabbrev]{cleveref}

\theoremstyle{plain}
\newtheorem{theorem}{Theorem}[section]

\theoremstyle{definition}
\newtheorem{definition}[theorem]{Definition}

\theoremstyle{remark}

\begin{document} 

\begin{abstract}
AI compilers aggressively rewrite computation graphs through normalization, lowering, and optimization, making it difficult to track the provenance of tensors and operators across compilation. Reliable provenance is essential for attaching platform-specific postprocessing, debugging compiler behavior, and validating transformations, yet existing solutions are either invasive or ad hoc under non-injective graph rewrites.

We present a lightweight, generative approach to provenance tracking based on observational semantics. Instead of propagating identifiers through compiler passes, we observe graph transformations and reason about provenance in terms of observable computational actions. We formalize this approach using a coalgebraic model and bisimulation, which preserves provenance even when intermediate nodes are eliminated. Furthermore, we implement this approach in a prototype AI compiler {\quartz}, demonstrating stable provenance across compilation pipelines with minimal engineering overhead.
\end{abstract}

\maketitle 

\section{Introduction}
\label{sec:intro}
AI compilers are a critical component of modern machine learning systems. They transform high-level models written in frameworks such as PyTorch~\cite{pytorch} or TensorFlow~\cite{tensorflow} into optimized, hardware-specific executables through a sequence of normalization, lowering, and optimization passes~\cite{tvm-acm, xla}. These transformations are essential for performance, but they also introduce a fundamental challenge: the loss of stable identity for computation graph nodes.

During compilation, operators may be fused, split, reordered, or eliminated entirely. As a result, a tensor or operator visible in the frontend model may not correspond to any single node in the optimized graph. This creates practical problems for developers and system integrators. For example, users often need to attach platform-specific postprocessing operators -— such as quantization, encoding, or proprietary transformations —- to selected tensors in the original model. These postprocessing operators are external to training, immutable, and must be correctly attached after compilation. Without reliable provenance, it becomes unclear which nodes in the optimized graph should receive these transformations.

Despite its importance, provenance tracking in AI compilers remains underdeveloped. Existing approaches typically fall into two categories. Tag-based methods propagate identifiers through compiler passes, but they require invasive modifications to each transformation. Post-hoc graph matching attempts to recover provenance by comparing graphs before and after optimization, but it is computationally expensive and unreliable under non-injective rewrites such as fusion. Neither approach scales to production compiler stacks.

More fundamentally, these approaches treat provenance as an auxiliary artifact to be recovered or maintained, rather than as a first-class correctness concern of compilation itself. As a result, they offer no principled way to determine whether a transformation preserves the information needed to explain how optimized code relates to its source. This gap raises a basic question: \emph{what does it mean for a compiler transformation to be correct with respect to provenance?}

Traditional compiler correctness is defined purely in terms of input–output behavior: an optimized graph is correct if it computes the same tensor values as the original. While this notion is sufficient for functional equivalence, it provides no guarantees about the preservation of structural or historical information during compilation. Intermediate operators that are fused away or rewritten remain semantically relevant for tasks such as postprocessing, debugging, and auditability, yet they are invisible under value-based correctness.

We address this question by shifting from a \emph{value-based} to an \emph{observation-based} notion of correctness. Rather than requiring compiler transformations to preserve only final tensor values, observational correctness demands the preservation of observable computational actions -- such as operator applications and rewrites -- and their evolution across compilation. By reasoning about these actions instead of purely structural graph artifacts, provenance can be preserved even when intermediate nodes are fused away or otherwise disappear during optimization.

Our key technical contribution is a coalgebraic formulation of observational correctness for AI compiler transformations. Coalgebra provides a natural framework for modeling state-based systems through their observable behavior. By modeling computation graphs as coalgebras and compilation passes as transformations between them, we define provenance preservation in terms of weak bisimulation over action-reflective semantics. Intuitively, two graphs are observationally equivalent if their observable computational actions agree, even if their internal structure differs.

Crucially, the coalgebraic formulation leads naturally to a generative implementation strategy. Rather than modifying compiler passes invasively, we automatically generate richer observational semantics from existing ones by reifying operator executions as observable events. This allows provenance information to be injected, propagated, and aggregated externally, without altering the compiler’s operational logic.

We implement the generative approach in a production AI compiler. Our system introduces provenance that integrates with PyTorch, TVM Relay IR, and quantization IR. Using lightweight compiler passes and decorators, provenance metadata is combined and updated across aggressive graph rewrites. This enables stable attachment of postprocessing operators and supports debugging and auditing workflows with minimal engineering overhead.

In summary, this work makes the following contributions:
\begin{itemize}
    \item We identify the limitations of value-based correctness for provenance tracking in AI compilers.
    \item We introduce observational correctness, formalized via coalgebra and weak bisimulation, as a principled foundation for provenance preservation.
    \item We present a generative instrumentation mechanism that realizes this theory in practice.
\end{itemize}

By bridging formal semantic reasoning and practical compiler engineering, this paper shows how observational semantics can make AI compilation more transparent, accountable, and developer-friendly.

The remainder of this paper is structured as follows. We first provide necessary background on coalgebraic concepts in Section~\ref{sec:background}. Section~\ref{sec:observational-correctness} details the idea of observational-correctness through a concrete example. 
Section~\ref{sec:implementation} describes the implementation within {\quartz} and evaluates its effectiveness. We discuss related work in Section~\ref{sec:related} and conclude with future directions in Section~\ref{sec:conclusion}.
\section{Mathematical Preliminaries}
\label{sec:background}

This section introduces the formal mathematical framework needed to define and reason about provenance in AI compilers. We start with a brief example motivating why existing approaches are not sufficient (\cref{subsec:core-challenge}), then introduce the minimal categorical concepts needed (\cref{subsec:math-preliminaries}), and finally show how coalgebra provides the right abstraction for tracking observable behavior through transformations (\cref{subsec:coalgebra-framework}).

\subsection{Challenge: From Syntactic to Semantic Identity}
\label{subsec:core-challenge}


Consider a PyTorch model where a user wants to specify custom operations for a convolutional operator \texttt{conv3}. During compilation, \texttt{conv3} may be fused with subsequent operations, decomposed for hardware support, or eliminated through constant folding. The resulting optimized graph bears little syntactic resemblance to the original, yet the {semantic action} performed by \texttt{conv3} -- its effect on the computation -- persists.

The fundamental problem is that compiler optimizations are designed to {preserve input-output values while altering structure}, but provenance requires {preserving the mapping between user intent and implementation}. 

\nop{Current approaches fail because they rely on either:
\begin{itemize}
    \item {Syntactic identity} (e.g., propagating tags), which is broken by structural changes like fusion,
    \item {Graph matching}, which is computationally expensive and fails under non-injective transformations
\end{itemize}
}
What is needed is a notion of {semantic identity} that survives structural changes. We capture this through  {observable computational actions}, explained in detail in \cref{sec:obs-correctness}, based on the coalgebra framework introduced in this section.

\subsection{Category Theory Basics}
\label{subsec:math-preliminaries}
To formalize observable behavior, we use category theory -- a mathematical framework for describing structure-preserving transformations.\footnote{Category theory is inherently structure-preserving because its notion of mapping—- functors -— is defined to preserve identities and composition, which encode all structural relationships within a category.} For our purposes, we need only three essential concepts:

\begin{definition}[Category]
    A {category} consists of:
    \begin{itemize}
        \item A collection of {objects} $\Ob(\mathbf{C})$
        \item For each pair of objects $X, Y \in \Ob(\mathbf{C})$, a set $\Hom_{\mathbf{C}}(X, Y)$ of {morphisms} from $X$ to $Y$
        \item An {identity morphism} $\mathrm{id}_X \in \Hom_{\mathbf{C}}(X, X)$ for each object $X$
        \item A {composition operation} $\circ : \Hom_{\mathbf{C}}(Y, Z) \times \Hom_{\mathbf{C}}(X, Y) \to \Hom_{\mathbf{C}}(X, Z)$ that is associative
    \end{itemize}
\end{definition}

\begin{definition}[Functor]
Given categories $\mathbf{C}$ and $\mathbf{D}$, a {functor} $F : \mathbf{C} \to \mathbf{D}$ consists of:
\begin{itemize}
    \item An object mapping: for each object $X \in \Ob(\mathbf{C})$, an object $F(X) \in \Ob(\mathbf{D})$
    \item A morphism mapping: for each morphism $f : X \to Y$ in $\mathbf{C}$, a morphism $F(f) : F(X) \to F(Y)$ in $\mathbf{D}$
\end{itemize}
satisfying:
\begin{enumerate}
    \item $F(\mathrm{id}_X) = \mathrm{id}_{F(X)}$ for all $X \in \Ob(\mathbf{C})$
    \item $F(g \circ f) = F(g) \circ F(f)$ for all composable morphisms $f, g$ in $\mathbf{C}$
\end{enumerate}
\end{definition}

\begin{definition}[Endofunctor]
    An {endofunctor} $F : \mathbf{C} \to \mathbf{C}$ maps a category to itself.
\end{definition}

Endofunctors play a central role in connecting category theory to stateful system
semantics. Intuitively, an endofunctor specifies the \emph{shape of observable
behavior} associated with objects of $\mathbf{C}$, abstracting away from how
those objects are constructed. This viewpoint naturally leads to coalgebras,
which model concrete systems as realizations of such behavioral interfaces.

\subsection{Coalgebra: A Framework for Observable Behavior}
\label{subsec:coalgebra-framework}

Coalgebra provides a natural mathematical model for systems whose behavior we can observe step-by-step---exactly what we need for tracking provenance through compilation.

\begin{definition}[$F$-coalgebra]
\label{def:coalgebra}
Given an endofunctor $F : \mathbf{C} \to \mathbf{C}$ on a category $\mathbf{C}$, an {$F$-coalgebra} is a pair $(S, \alpha : S \to F(S))$ where:
\begin{itemize}
    \item $S \in \Ob(\mathbf{C})$ is the {state space} or carrier object
    \item $\alpha : S \to F(S)$ is a morphism in $\mathbf{C}$ called the {transition structure} and captures observations and continuations from each state.
\end{itemize}
\end{definition}
For example, a convolution operation might be modeled as a state where $\alpha$ observes ``Conv'' and continues to the next operation's state.
This captures exactly the {observable action} we need to track for provenance: what operation is performed and what happens next.

\begin{definition}[Bisimulation]
A relation $R \subseteq S \times T$ between states of two coalgebras $(S,\alpha)$ and $(T,\beta)$ is a {bisimulation} if there exist $\rho : R \to F(R)$ making the projections $\pi_1 : R \to S$ and $\pi_2 : R \to T$ into coalgebra morphisms, i.e. for $i \in \{1,2\}$, 
$
\alpha \circ \pi_i = F(\pi_i) \circ \rho.
$
\end{definition}

\begin{figure}[h]
\centering
\begin{tikzcd}[row sep=large, column sep=huge]
& R \arrow[dl, "\pi_1"'] \arrow[dr, "\pi_2"] \arrow[d, "\rho" description] & \\
S \arrow[d, "\alpha"'] & F(R) \arrow[dl, "F(\pi_1)"] \arrow[dr, "F(\pi_2)"'] & T \arrow[d, "\beta"] \\
F(S) & & F(T)
\end{tikzcd}
\caption{Diagram of bisimulation: $R$ relates states of $S$ and $T$,
and $\rho$ ensures the coalgebra structures are compatible via projections.}
\label{fig:f-bisimulation}
\end{figure}
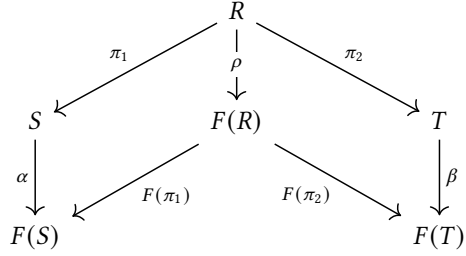

Bisimulation gives us a precise way to relate two differently structured graphs: when two states are bisimilar, they are {observationally indistinguishable} -- the foundation of our provenance guarantees. \Cref{fig:f-bisimulation} shows a high-level diagram to illustrate it.

\begin{theorem}[Coinduction Principle]
    \label{thm:coinduction}
If a relation $R$ between states of coalgebras is a bisimulation, then all related states are observationally equivalent. This allows us to prove equivalence by checking stepwise compatibility rather than analyzing entire executions.
\end{theorem}

The formal definition of coinduction principle can be found in appendix. In the next section, we use the framework of coalgebra to define {observational correctness} -- a formal criterion requiring that compiler transformations preserve not just input-output values but also the observable action history needed for provenance tracking.

\section{Observational Correctness}
\label{sec:observational-correctness}
In this section, we begin by reviewing the conventional correctness condition for fused operators in AI compilers, which adopts a value-based semantics. While suitable for verifying functional equivalence, this criterion is insufficient for provenance tracking, where we must also account for the transformation history of operators throughout successive rewritings. In particular, intermediate operators that are fused away remain unmaterialized -- and thus unobservable -- under purely value-based reasoning. 

To address this gap, we introduce the notion of \emph{observational correctness}, grounded in the observation of program actions throughout compilation. We then establish how this new condition relates to the standard value-based criterion using techniques from generative programming.

\subsection{The Limitation of Value-based Correctness}
\label{sec:value-correctness}
We model computation graphs as coalgebras whose transitions are labeled by
{operator symbols}. Let $X$ be a set of computation states,
$T$ the set of input tensors,
$V$ the set of tensor values,
and $\Sigma$ a set of observable operator labels.
Let $\mathbf{1} = \{\mathtt{stop}\}$ denote the singleton set representing
the termination.

The continuation functor is defined as:
\begin{equation}
\label{eq:cont-functor}
F(X) = \mathcal{P}\bigl((\Sigma \times (T \to (V \to X))) + \mathbf{1}\bigr).
\end{equation}

An $F$-coalgebra $(X,\gamma)$ assigns to each state $x \in X$ a set of
{computational continuations}, each consisting of
\begin{itemize}
    \item a tuple $(\sigma, \kappa)$ with a label $\sigma \in \Sigma$ and a continuation $\kappa: T\to (V \to X)$ mapping an input tensor $t \in T$ and the resulting value $v \in V$
        to the next state,
    \item or a distinguished termination observation $\mathtt{stop} \in \textbf{1}$.
\end{itemize}

\paragraph{Example.}
Let $X = \{s_0, s_1, s_2\}$, and define $\gamma : X \to F(X)$ by:
\begin{itemize}
    \item At $s_0$:
    $
    \gamma(s_0)
    =
    \bigl\{
    (\texttt{Conv}, \lambda t.\lambda v.\, s_1)
    \bigr\},
    \quad
    v = \llbracket \texttt{Conv} \rrbracket(t).
    $

    \item At $s_1$:
    $
    \gamma(s_1)
    =
    \bigl\{
    (\texttt{ReLU}, \lambda t.\lambda v.\, s_2)
    \bigr\},
    \quad
    v = \llbracket \texttt{ReLU} \rrbracket(v).
    $

    \item At $s_2$:
    $
    \gamma(s_2) = \{\mathtt{stop}\}.
    $
\end{itemize}

This coalgebra denotes a computation that applies $\texttt{Conv}$ followed by
$\texttt{ReLU}$ to an input tensor and then terminates, yielding a final tensor
value. From the entry state $s_0$, the denotation of the coalgebra coincides
with the composite function:
$$
\llbracket \texttt{ReLU} \rrbracket \circ \llbracket \texttt{Conv} \rrbracket : T \to V.
$$

\paragraph{Operator Fusion as Coalgebra Transformation.}
Fusion is {not} a simple coalgebra homomorphism, as it can alter the {carrier set}, {transition labels}, and continuations.
Instead, we model fusion as a \emph{coalgebra transformation} that rewrites an unfused computation graph
$(X,\gamma)$ into $(X',\gamma')$.

Continue with the running example. We define the fused coalgebra $(X',\gamma')$ by taking
\[
X' = \{s_0, s_2\}
\]
and
\[
\gamma' : X' \to \mathcal{P}\bigl((\Sigma \times (T \to (V \to X'))) + \mathbf{1}\bigr)
\]
as follows:
\[
\gamma'(s_0)
=
\bigl\{
(\texttt{ConvReLU}, \lambda t.\lambda v.\, s_2)
\bigr\},
\quad
\gamma'(s_2) = \{\mathtt{stop}\}.
\]
The fused operator satisfies
\[
\llbracket \texttt{ConvReLU} \rrbracket(t)
=
\llbracket \texttt{ReLU} \rrbracket
\circ \llbracket \texttt{Conv} \rrbracket(t),
\]
thereby composing the computational effects of the two original steps.
The intermediate state $s_1$ is eliminated, yielding a single-step transition.

\paragraph{Value-based Correctness.}
The correctness of fusion is stated and proved semantically, using a purely value-based
interpretation of programs.
Let $(\nu F,\zeta)$ be the final $F$-coalgebra.
By finality, there exist unique coalgebra morphisms
\[
\llbracket \cdot \rrbracket_\gamma : X \to \nu F,
\qquad
\llbracket \cdot \rrbracket_{\gamma'} : X' \to \nu F,
\]
which assign to each state its observable behavior.

Under the standard \emph{black-box semantics} adopted by AI compilers, programs
are compared solely by the values they compute at designated \emph{entry
states} (e.g., the initial state of a computation graph), with all intermediate
states treated as semantically irrelevant.
In this setting, fusion is deemed correct if the denotation of the initial state
$s_0$ is preserved:
\[
\llbracket s_0 \rrbracket_\gamma
=
\llbracket s_0 \rrbracket_{\gamma'}.
\]
This equality expresses functional equivalence of the unfused and fused programs
and constitutes the conventional correctness criterion for compiler
optimizations.

The proof proceeds by coinduction. After one unfolding step, both coalgebras
expose the same composite computation as a single denotational value, and the
remaining structure contributes identically to this value. Consequently, the unfused computation
\[
s_0 \xrightarrow{\texttt{Conv}} s_1
\xrightarrow{\texttt{ReLU}} s_2 \xrightarrow{} \mathtt{stop}
\]
and the fused computation
\[
s_0 \xrightarrow{\texttt{ConvReLU}} s_2 \xrightarrow{} \mathtt{stop}
\]
induce the same element of the final coalgebra when evaluated from the entry state
$s_0$.

\paragraph{Limitation for Provenance Tracking.}
The correctness result established above is formulated with respect to a
semantics that compares programs only at entry states and equates
them by the tensor values they compute. This reflects the conventional
{black-box} view of AI compilers.

One might attempt to refine this semantics by reinterpreting the transition
labels of the coalgebra as {observable operator} executions, thereby shifting
from a value-based to an observation-based reading. However, \emph{without modifying
the underlying continuation functor, such a reinterpretation does not expose
additional provenance information}. The reason is structural: the functor $F$
assigns meaning to a computation solely through its unfolding from an entry
state into the final coalgebra, and this unfolding yields a single denotation
that subsumes all intermediate states.

\subsection{Observational Correctness via Action Traces} 
\label{sec:obs-correctness}

In fused computation graphs, intermediate tensor values are no longer
materialized and therefore cannot be observed directly. As a consequence,
correctness cannot be formulated in terms of value equality at intermediate
states. What remains observable, however, is the structure of computation as
revealed by the sequence of operator executions induced by compilation and
rewriting. To capture this information, we shift the semantic focus from
intermediate values to \emph{computational actions}, and relate a fused operator
to the trace of actions it subsumes.

\paragraph{Action Reflection.}
Given an $F$-coalgebra $(X,\gamma)$, its \emph{action-reflected $F^\sharp$-coalgebra}
$(X^\sharp,\gamma^\sharp)$ has carrier
\[
X^\sharp \;\triangleq\; X \;\cup\; (X \times \Sigma),
\]
where $x \in X$ are control states and $(x,\sigma)$ are explicit action states
representing the execution of $\sigma$ at $x$.
$F^\sharp$ is an \emph{action-reflecting functor} that records operator executions as first-class observable events:
\[
F^\sharp(X) \;\triangleq\;
\mathcal{P}\Bigl(
(X \times \Sigma)
\;\cup\;
\bigl(\Sigma \times (T \to (V \to X))\bigr)
\;\cup\;
\mathbf{1}
\Bigr).
\]

Compared to $F$, 
the additional $X \times \Sigma$ term ensures that operator 
executions are not lost in the final semantics. The fused actions remain observable
as part of an action trace.

For each transition $(\sigma,k) \in \gamma(x)$, the reflected coalgebra exposes
\[
x \;\longrightarrow\; (x,\sigma) \;\longrightarrow\; k(\cdot),
\]
reifying operator executions as first-class observable events. Coalgebraically,
this turns edge-labeled transitions into explicit intermediate states, ensuring
all actions are visible.

\paragraph{Example.}
For the unfused computation graph
\[
s_0 \xrightarrow{\texttt{Conv}} s_1 \xrightarrow{\texttt{ReLU}} s_2,
\]
action reflection produces
\[
s_0 \;\to\; (s_0,\texttt{Conv}) \;\to\; s_1 \;\to\; (s_1,\texttt{ReLU}) \;\to\; s_2,
\]
where both \texttt{Conv} and \texttt{ReLU} appear as explicit intermediate states.
The fused computation
\[
s_0 \xrightarrow{\texttt{ConvReLU}} s_2
\]
yields
\[
s_0 \;\to\; (s_0,\texttt{ConvReLU}) \;\to\; s_2,
\]
so although the intermediate tensor $s_1$ is eliminated, the generated operator
remains observable.

\paragraph{Action Abstraction.}
Action reflection highlights an asymmetry: unfused executions produce sequences
of primitive actions, while fused executions produce single generated actions.
We introduce an \emph{action abstraction relation}
\[
A \;\subseteq\; \Sigma^{*} \times \Sigma',
\]
relating sequences of primitive operators to generated operators. For instance,
\[
(\texttt{Conv},\texttt{ReLU}) \;A\; \texttt{ConvReLU}.
\]
At the level of action states, this induces a correspondence
\[
\bigl((s_0,\texttt{Conv}),(s_1,\texttt{ReLU})\bigr) \;\sim\; (s_0,\texttt{ConvReLU}),
\]
allowing generated actions to summarize sequences of primitive actions.

\paragraph{Observational Correctness.}
We define a relation
\[
R \;\subseteq\; X^\sharp \times X'^\sharp
\]
relating states and action states of unfused and fused graphs:

\begin{itemize}
    \item $(s_0, s_0) \in R$ (control state),
    \item $((s_0,\texttt{Conv}), (s_0,\texttt{ConvReLU})) \in R$, \\
          $((s_1,\texttt{ReLU}), (s_0,\texttt{ConvReLU})) \in R$ (action states),
    \item $(s_2, s_2) \in R$ (termination).
\end{itemize}

$R$ is an $F^\sharp$-bisimulation: each related pair of states has transitions
whose successors are also related, respecting the action abstraction $A$. For
example, $\gamma^\sharp(s_0) = \{(s_0,\texttt{Conv})\}$ and
$\gamma'^\sharp(s_0) = \{(s_0,\texttt{ConvReLU})\}$, and
$((s_0,\texttt{Conv}),(s_0,\texttt{ConvReLU})) \in R$ ensures the successors
are related; the argument extends to all control, action, and termination states.

By coinduction, all related states are mapped to the same element of the final
$F^\sharp$-coalgebra:
\[
\llbracket s_0 \rrbracket_{\gamma^\sharp} =
\llbracket s_0 \rrbracket_{\gamma'^\sharp},
\llbracket (s_1,\texttt{ReLU}) \rrbracket_{\gamma^\sharp} =
\llbracket (s_0,\texttt{ConvReLU}) \rrbracket_{\gamma'^\sharp}.
\]

\paragraph{Implications for Fusion and Provenance.}
Fusion is observationally correct with respect to action-based semantics. While
intermediate tensor values and control states may be eliminated, the
corresponding computational actions remain observable and comparable up to
abstraction. This establishes a principled foundation for provenance tracking:
generative compilation does not conflict with provenance, but requires a shift
from value-based to action-based semantics, which is naturally formalized and
coinductively verified via the coalgebraic framework.

\subsection{Bridging Value-based and Observational Correctness Generatively}
The distinction between value-based and observational correctness introduced
above is semantic rather than algorithmic: both notions reason about the same
underlying compilation process, but differ in what aspects of that process are
made observable. In this subsection, we show that the transition from
value-based correctness (Section~\ref{sec:value-correctness}) to
action-based observational correctness (Section~\ref{sec:obs-correctness}) can
be understood as a form of \emph{generative programming}, in which the original
semantic functor is systematically transformed to expose additional structure.

\paragraph{Functorial Refinement by Generation.}
Recall that value-based correctness is formulated over the $F$ functor:
\[
F(X) = \mathcal{P}\bigl((\Sigma \times (T \to (V \to X))) + \mathbf{1}\bigr),
\]
while observational correctness relies on the $F^{\#}$ functor:
\[
F^\sharp(X) =
\mathcal{P}\Bigl(
(X \times \Sigma)
\;\cup\;
(\Sigma \times (T \to (V \to X)))
\;\cup\;
\mathbf{1}
\Bigr).
\]
Crucially, $F^\sharp$ is not an unrelated semantic model: it is obtained from
$F$ by a systematic enrichment that reifies operator executions as explicit
semantic events. From a generative perspective, $F^\sharp$ can be seen as a
\emph{generated variant} of $F$, produced by adding a single observation layer
while leaving the underlying continuation structure unchanged.

\paragraph{Minimal Change Principle.}
From the standpoint of compiler implementation, the continuation structure
$\Sigma \times (T \to (V \to X))$ already embodies all information required to
construct action traces. Generative programming exploits this fact by
automatically transforming each transition
\[
(\sigma,k) : X \to \Sigma \times (T \to (V \to X))
\]
into a two-stage transition
\[
x \;\mapsto\; (x,\sigma) \;\mapsto\; k(\cdot),
\]
without modifying the original evaluation logic encoded in $k$. In particular,
the computation of tensor values, scheduling decisions, and optimization
passes remain unchanged; only the semantic interface through which behavior is
observed is refined.

This separation of concerns is essential: provenance tracking is introduced not
by altering the compiler’s operational behavior, but by \emph{generating} a
richer semantic model from an existing one.

\paragraph{Generative Semantics as a Functor Transformer.}
Formally, action reflection can be viewed as a functor transformer
\[
(-)^\sharp : \mathbf{Coalg}(F) \;\to\; \mathbf{Coalg}(F^\sharp),
\]
which maps an $F$-coalgebra $(X,\gamma)$ to an $F^\sharp$-coalgebra
$(X^\sharp,\gamma^\sharp)$ by a uniform syntactic construction. Coalgebra morphisms are preserved, 
and value-based correctness results lift to the action-reflected setting.

Hence, existing correctness proofs for compiler optimizations -- such
as fusion -- can be reused with minimal adaptation. What changes is not the proof
technique (coinduction), but the observational granularity of the semantic
domain.

\paragraph{From Values to Actions, and Back.}
The relationship between value-based and observational correctness is therefore
one of \emph{semantic projection}. Action-based semantics refines value-based
semantics by retaining additional structure; value-based correctness is
recovered by forgetting action states and projecting action traces back to
their induced input–output behavior. In this sense, $F^\sharp$ is a conservative
extension of $F$: it supports strictly finer observations while preserving all
value-based equivalences.

\paragraph{Implications.}
The generative correspondence between $F$ and $F^\sharp$ suggests a lightweight
implementation strategy for provenance tracking. Rather than redesigning compiler 
passes or introducing ad hoc instrumentation,
one can exploit the functorial transformation from $F$ to $F^\sharp$ as a
\emph{semantic code generator}, lifting a value-based continuation semantics
into an action-reflecting one. 

Concretely, the generative correspondence suggests that provenance tracking 
can be implemented as a systematic transformation over the compiler’s IR or
execution graph: each operator transition is generatively wrapped to emit an
explicit action event before delegating to the original continuation. No
changes are required to the evaluation of tensor values or to optimization
passes such as fusion; these passes remain correct by construction, while their
effects become observable at the action level.
\section{\quartz}
\label{sec:implementation}
We implement observational provenance tracking in a prototype compiler {\quartz}. Instead of propagating identifiers through the IR,
we instrument compilation stages to expose observable computational actions and
relate them across transformations. This realizes the coalgebraic semantics
introduced in \cref{sec:observational-correctness} in a lightweight AI compiler.

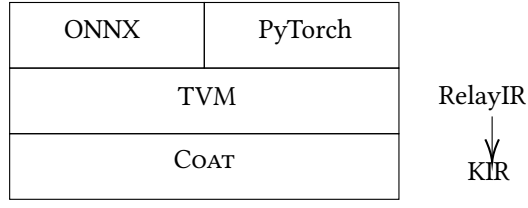
\begin{figure}
    \centering
    \resizebox{0.5\columnwidth}{!}{\begin{tikzpicture}[x=0.75pt,y=0.75pt,yscale=-1,xscale=1]

\draw   (53,72) -- (147,72) -- (147,103.67) -- (53,103.67) -- cycle ;
\draw   (147,72) -- (241,72) -- (241,103.67) -- (147,103.67) -- cycle ;
\draw   (53,103.67) -- (241,103.67) -- (241,135.33) -- (53,135.33) -- cycle ;
\draw   (53,135.33) -- (241,135.33) -- (241,167) -- (53,167) -- cycle ;
\draw    (286,127) -- (286,147) ;
\draw [shift={(286,149)}, rotate = 270] [color={rgb, 255:red, 0; green, 0; blue, 0 }  ][line width=0.75]    (10.93,-3.29) .. controls (6.95,-1.4) and (3.31,-0.3) .. (0,0) .. controls (3.31,0.3) and (6.95,1.4) .. (10.93,3.29)   ;

\draw (78,78.89) node [anchor=north west][inner sep=0.75pt]   [align=left] {ONNX};
\draw (172,78.89) node [anchor=north west][inner sep=0.75pt]   [align=left] {PyTorch};
\draw (133,111.44) node [anchor=north west][inner sep=0.75pt]   [align=left] {TVM};
\draw (131,142.11) node [anchor=north west][inner sep=0.75pt]   [align=left] {{\qatc}};
\draw (259,110.44) node [anchor=north west][inner sep=0.75pt]   [align=left] {RelayIR};
\draw (273,147.44) node [anchor=north west][inner sep=0.75pt]   [align=left] {KIR};

\end{tikzpicture}}
    \caption{{\quartz} Software Architecture and IR Flow}
    \label{fig:quartz-arch}
\end{figure}

\subsection{Compiler Context}
{\quartz} is a multi-layer AI compiler that separates frontend normalization,
semantic graph rewriting, quantization, and hardware-specific lowering
(\cref{fig:quartz-arch}). Models from PyTorch, TensorFlow, and ONNX are
lowered into TVM Relay IR, where most semantics-preserving but structurally
non-injective rewrites -- such as fusion and normalization -- occur.
The Relay graph is then translated by {\qatc} into the {\quartz} KIR Model with
quantized operators, after which backend lowering preserves graph structure.

Because later stages do not alter graph topology, provenance tracking focuses on
the transformation pipeline from frontend IRs through Relay and into the KIR
Model. This boundary cleanly isolates the stages where observational equivalence
must be preserved.

\subsection{Observation-Aware IR Flow}
We illustrate the provenance flow using PyTorch as the frontend.
PyTorch models are exported via TorchScript, which normalizes execution into
SSA-form graphs of \texttt{ATen} operators. While the original module hierarchy
is erased, TorchScript assigns deterministic operator identifiers, providing a
stable anchor for user-defined semantics.

Relay IR represents the primary site of semantic graph rewriting. Operators may
be reordered, fused, or eliminated, destroying syntactic correspondence with
the frontend graph. To preserve observational correctness, we attach provenance
metadata to Relay operators using spans derived from frontend operator anchors.
These spans record observable computational actions rather than node identity.

During Relay transformations, spans are propagated according to observable
behavior: fused operators aggregate the spans of their constituents, while
splitting operations copy spans to all results. This produces an action
trace that remains stable even under non-injective rewrites.

\subsection{Materialization in the KIR Model}
The {\qatc} layer translates Relay operators into quantized operators in the KIR
Model. At this stage,
aggregated action traces are consolidated, and user-specific actions like attaching postprocessing flows are applied to the transformed graph, fixing the granularity of observable actions at the
implementation level. 

Once operators are materialized in the KIR Model, later compilation stages no
longer modify graph structure. Provenance tracking is therefore complete:
Each quantized operator is associated with an action-level lineage that
connects frontend semantics to hardware-oriented execution. Depending on the requirements, the provenance information can hence be removed in later stages. 

\nop{
This section describes how the action-based observational correctness semantics
introduced before can be reified in a lightweight
provenance tracking system. Rather than modifying the core compilation logic,
we show how existing compilation stages in {\quartz}~\cite{quartz} can be instrumented to expose
and relate observable computational actions, even in the presence of aggressive
graph optimizations.

\subsection{Architectural Context}
\Cref{fig:quartz-arch} situates our provenance tracking mechanism within the 
overall {\quartz} system architecture. {\quartz} is a multi-layer compilation 
stack that cleanly separates frontend normalization, semantic
graph rewriting, operator quantization, legalization, and hardware-specific optimizations. This
separation provides natural insertion points for implementing observational
correctness without interfering with existing compilation passes.

At the top of the software stack, {\quartz} supports multiple machine learning frontends,
including PyTorch~\cite{pytorch}, TensorFlow~\cite{tensorflow}, and ONNX~\cite{onnx}. Models originating from these
frameworks are translated into TVM's Relay IR~\cite{tvm, tvm-acm}, which serves as the unified
graph-level representation. The TVM layer is responsible for early 
graph transformations, including normalization, operator fusion, and
other semantics-preserving rewritings that may substantially alter graph
topology.

Below TVM, {\quartz} introduces the {\qatc} layer, which bridges Relay IR and
{\quartz} KIR Model with quantized operators. {\qatc} translates Relay IR nodes into 
{\quartz}-supported operator nodes, enforcing operator availability,
input shape constraints, and representation invariants required by the target
NPU backends. While {\qatc} does not introduce new semantic rewritings, it plays a
critical role in fixing the granularity at which operators become observable to
the rest of the system.

The lower portion of the stack comprises the NPU compiler that translates KIR Model into QConfigIR and a collection of
backend code generators that generate OVM for different hardware backends (e.g., OG12B, AOP2, XNPU). 
These stages focus on hardware-aware
lowering, including memory allocation, scheduling, and hardware-specific optimizations. The graph structure remains unchanged.

\paragraph{Implication.}
For provenance tracking, we are mostly concerned with how a node in PyTorch -- or other frontend frameworks -- is transformed through normalization, optimization, and other transformations. Since QConfigIR and OVM do not have alter graph structures, hence we only focus on the transformation from PyTorch to Relay IR to KIR Model for provenance tracking.

\begin{figure}
    \centering
    \begin{subfigure}[b]{0.8\columnwidth}
        \resizebox{\textwidth}{!}{\begin{tikzpicture}[x=0.75pt,y=0.75pt,yscale=-1,xscale=1]

\draw   (53,72) -- (147,72) -- (147,103.67) -- (53,103.67) -- cycle ;
\draw   (147,72) -- (241,72) -- (241,103.67) -- (147,103.67) -- cycle ;
\draw   (53,103.67) -- (241,103.67) -- (241,135.33) -- (53,135.33) -- cycle ;
\draw   (53,135.33) -- (241,135.33) -- (241,167) -- (53,167) -- cycle ;
\draw    (286,127) -- (286,147) ;
\draw [shift={(286,149)}, rotate = 270] [color={rgb, 255:red, 0; green, 0; blue, 0 }  ][line width=0.75]    (10.93,-3.29) .. controls (6.95,-1.4) and (3.31,-0.3) .. (0,0) .. controls (3.31,0.3) and (6.95,1.4) .. (10.93,3.29)   ;

\draw (78,78.89) node [anchor=north west][inner sep=0.75pt]   [align=left] {ONNX};
\draw (172,78.89) node [anchor=north west][inner sep=0.75pt]   [align=left] {PyTorch};
\draw (133,111.44) node [anchor=north west][inner sep=0.75pt]   [align=left] {TVM};
\draw (131,142.11) node [anchor=north west][inner sep=0.75pt]   [align=left] {{\qatc}};
\draw (259,110.44) node [anchor=north west][inner sep=0.75pt]   [align=left] {RelayIR};
\draw (273,147.44) node [anchor=north west][inner sep=0.75pt]   [align=left] {KIR};

\end{tikzpicture}}
        \caption{Software Stack}
    \label{fig:quartz-arch}
    \end{subfigure}
    \hfill
    \begin{subfigure}[b]{0.15\columnwidth}
        \resizebox{\textwidth}{!}{\begin{tikzpicture}[x=0.75pt,y=0.75pt,yscale=-1,xscale=1]

\draw    (108.43,54.29) -- (108.43,70.29) ;
\draw [shift={(108.43,72.29)}, rotate = 270] [color={rgb, 255:red, 0; green, 0; blue, 0 }  ][line width=0.75]    (10.93,-3.29) .. controls (6.95,-1.4) and (3.31,-0.3) .. (0,0) .. controls (3.31,0.3) and (6.95,1.4) .. (10.93,3.29)   ;
\draw    (108.43,95.29) -- (108.43,111.29) ;
\draw [shift={(108.43,113.29)}, rotate = 270] [color={rgb, 255:red, 0; green, 0; blue, 0 }  ][line width=0.75]    (10.93,-3.29) .. controls (6.95,-1.4) and (3.31,-0.3) .. (0,0) .. controls (3.31,0.3) and (6.95,1.4) .. (10.93,3.29)   ;
\draw    (108.43,133.29) -- (108.43,149.29) ;
\draw [shift={(108.43,151.29)}, rotate = 270] [color={rgb, 255:red, 0; green, 0; blue, 0 }  ][line width=0.75]    (10.93,-3.29) .. controls (6.95,-1.4) and (3.31,-0.3) .. (0,0) .. controls (3.31,0.3) and (6.95,1.4) .. (10.93,3.29)   ;

\draw (83.67,36) node [anchor=north west][inner sep=0.75pt]   [align=left] {RelayIR};
\draw (81.67,75) node [anchor=north west][inner sep=0.75pt]   [align=left] {QDModel};
\draw (92.67,152) node [anchor=north west][inner sep=0.75pt]   [align=left] {OVM};
\draw (78.67,115) node [anchor=north west][inner sep=0.75pt]   [align=left] {QConfigIR};

\end{tikzpicture}}
        \caption{IRs}
    \label{fig:ir}
    \end{subfigure}
    \caption{{\quartz} Architecture and IR Flow}
    \label{fig:quartz-arch-ir}
\end{figure}

\subsection{Observation-Aware IR Flow}
{\quartz} supports different frameworks in the frontend, here we choose PyTorch for illustration. \Cref{alg:annotated-pytorch} shows an annotated PyTorch example. Below, we describe how this PyTorch example is translated through each pipeline in {\quartz}, focusing on the characteristics of each IR and observable information.






\paragraph{PyTorch and TorchScript (Frontend)}
PyTorch models are defined as \texttt{torch.nn.Module} subclasses and executed imperatively with dynamic computation graphs. Users can export models via \texttt{@torch.jit.trace} and annotate methods with \texttt{@torch.jit.export}, as shown in \cref{alg:annotated-pytorch}, creating custom entry points in the TorchScript graph, which is normalized to SSA form. 

In TorchScript, PyTorch module operations are compiled to \texttt{ATen} operators and modulel hierarchy is not preserved, making it difficult to infer the initial structure information from the normalized graph. TorchScript assigns unique, deterministic operator identifiers, suffixed sequentially (e.g., \texttt{aten::\_\_convolution\_\_3}) in the order operations are encountered. This provides a stable mapping from user-defined names to \texttt{ATen} operators across repeated executions.

Our approach leverages this implicit provenance: user-defined names act as anchors to associate post-processing logic with the corresponding \texttt{ATen} operators—without modifying PyTorch’s tracing or scripting infrastructure.

\paragraph{Relay IR (TVM)}
Relay IR represents programs as functional dataflow graphs and is the workhorse of semantic graph rewriting. Unlike TorchScript, which is specific to PyTorch, Relay IR is a framework-agnostic intermediate representation that can encode models from multiple frontends, including ONNX, TensorFlow, and others. This model-independence makes Relay IR an ideal stage for implementing semantic optimizations and provenance tracking in a unified manner.

In our running example, operators imported from TorchScript -- such as convolutions or elementwise operations -- are expressed as Relay nodes connected by dataflow edges. At this stage, the original Python module hierarchy is lost; the graph encodes only computation and dependencies. Operators may be reordered, fused, or split, reflecting the transformations performed by Relay passes.

To preserve observational correctness, we inject provenance identifiers derived from user-defined module names into the span of each Relay operator, in an added compilation pass, like \texttt{Span(module=conv3)}. Spans serve as metadata recording the original computational source. During fusion, spans of all constituent operators are aggregated into an ordered trace, capturing an action-level summary of the transformation. When an operator is split, its span is copied to each resulting node, ensuring provenance is propagated consistently. Through this mechanism, every Relay operator carries a precise record of its computational lineage, maintaining a stable mapping from the original TorchScript operators to the transformed Relay graph and forming the foundation of our provenance tracking framework.

\paragraph{KIR Model ({\qatc})}
The {\qatc} layer translates Relay IR into the {\quartz} KIR Model by mapping Relay operators to quantized operators supported by the target NPUs. While operator fusion and span aggregation occur in Relay IR, the KIR Model layer \emph{materializes} these transformations: it generates new operators and assigns new span information that consolidates the provenance of all constituent operations. 

For example, a convolution operator in Relay IR with span information \texttt{Span(module=conv3)}  may become a fused operator \texttt{ConvRelu2} in the KIR Model -- possibly with a different suffix reflecting the new graph ordering -- reflecting the combined lineage.

At this stage, the granularity of observable actions is fixed, corresponding to the quantized, implementation-level operators. Action traces from the original Relay IR are reinterpreted and associated with the new operators, ensuring that the consolidated provenance accurately reflects both the computational lineage and any graph modifications, such as the attachment of post-processing nodes. Once we reach the KIR Model, there is no need for operator-level provenance tracking, as later IR flows do not modify graph structure.

\nop{
\subsection{Generative Provenance Instrumentation}
The generative provenance instrumentation system implements action-aware provenance tracking by both (i) injecting module-level provenance into Relay IR operators and (ii) leveraging this information to generate materialized nodes in the KIR Model.

During compilation, each Relay operator is annotated with a span containing its originating TorchScript module name information. As operators are transformed -- fused, split, or reordered -- these spans are propagated and aggregated, producing a deterministic, action-reflective record of computation. In {\qatc}, the spans are then consolidated to reflect fused and quantized operators, establishing a stable, implementation-level mapping from high-level module names to hardware-oriented execution.

In practice, provenance injection is realized as a TVM compiler pass. When multiple operators are fused in Relay IR, their spans are combined into an ordered trace, while splitting operations copy the original span to all resulting nodes. This ensures that every transformed operator retains a complete record of its computational lineage.

Figure~\ref{fig:an-tvm-pass} illustrates an example of an annotated compiler optimization pass. The \texttt{@inject\_span} decorator automates span management: it extracts fused call spans from the pre-transformation graph, executes the pass’s transformation logic, and sets updated spans on the resulting expressions. The \texttt{ATenIRSpanManager} encapsulates the underlying logic for constructing, retrieving, and fusing spans, including handling duplicates and generating deterministic fused identifiers for composite operators.

\texttt{ATenIRSpanManager} embodies generative programming principles by automatically producing and maintaining provenance metadata. Its responsibilities include:
\begin{itemize}
    \item Injecting module names into spans.
    \item Aggregating spans during fusion.
    \item Copying spans during splitting.
    \item Extracting module names from spans for downstream passes or post-processing.
\end{itemize}
By automating these actions, \texttt{ATenIRSpanManager} ensures that provenance metadata is generated, fused, and propagated consistently across transformations.

The \texttt{@inject\_span} decorator operationalizes this generative approach for individual TVM passes. For each annotated pass, such as \texttt{ReshapeToFlatten}, it:
\begin{enumerate}
    \item Extracts fused call spans from the pre-transformation graph.
    \item Executes the pass’s transformation logic.
    \item Injects or updates spans on the resulting operators.
\end{enumerate}
This separation allows each pass to focus solely on graph rewriting, while the generative instrumentation maintains a complete, action-reflective record of computational lineage.

Finally, the spans generated and propagated through Relay IR are used in {\qatc} to materialize user-specified graph transformations in the transformed operators. {\qatc} layer correctly identifies which operators in the transformed graph correspond to user-specified nodes and performs specified actions, like attaching the specified postprocessing subgraph to the transformed nodes. 
}
}

\section{Evaluation}
\label{sec:evaluation}

We evaluate the effectiveness of our coalgebraic provenance framework by measuring its {instrumentation overhead}, both in terms of \emph{engineering efforts} and \emph{performance}. We show that our approach is light-weight with little changes to the current flow and negligible instrumentation overhead. 

\paragraph{Experimental Setup.}
Experiments are conducted using a 12th Gen Intel(R) i9-12900 server, featuring 16 cores and 24 hardware threads. The operating system is Ubuntu 22.04. 

For the benchmark, we select modified MNIST~\cite{mnist} and MobilenetV2-0.25~\cite{mbv2} as our benchmark, which are the most relevant to our internal applications. Other test results are over proprietary models, hence omitted. 



\paragraph{Engineering Effort.}
Provenance tracking is realized as an auxiliary pass that
annotates spans during compilation and a macro that injects span transformation to each pass, shown in \cref{fig:an-tvm-pass}. In practice, this adds minimal
engineering overhead and does not affect compilation correctness or optimization
behavior.

\begin{algorithm}
\caption{Annotated TVM Compiler Optimization Pass}
\label{fig:an-tvm-pass}
\begin{algorithmic}[1]
\STATE \textbf{class} \textsc{ConvFusion}
\STATE \hspace{1em}\textbf{initialize} pass configuration

\vspace{0.5em}
\STATE \hspace{1em}\COMMENT{Callback invoked on each rewrite}
\STATE \hspace{1em}\textbf{@inject\_span}
\STATE \hspace{1em}\textbf{function} \textsc{Callback}$(pre, post, node\_map)$
\STATE \hspace{2em}\COMMENT{Attach provenance span to rewritten nodes}
\STATE \hspace{2em}\COMMENT{Update node correspondence map}
\STATE \hspace{2em}$\cdots$
\STATE \hspace{1em}\textbf{end function}

\end{algorithmic}
\end{algorithm}

\paragraph{Performance Overhead.} 
We are interested in how well our compiler optimizes the model and the computation overhead of the provenance system. Hence, we measure the number of operators for models in our benchmark both before and after applying NPU optimizations, as well as the execution time with and without the provenance enabled. 

Table~\ref{tab:coverage_overhead} summarizes the results. The reduction rate for both models are 67\% and 52\% respectively, and the computation overhead of 
the provenance system is 5\% and 12\%. The overall end-to-end latency increases by less than 1s, which is negligible in practical compilers. This validates the core theoretical claim: observational correctness via action traces provides a robust semantic basis for lineage that survives non-injective rewrites with negligible overhead.

\begin{table}[t]
\centering
\caption{Provenance coverage and runtime overhead for a suite of model compilation tasks. Overhead is measured relative to the uninstrumented baseline.}
\label{tab:coverage_overhead}
\begin{tabular}{lcccc}
\toprule
\textbf{Model} & \textbf{Original} & \textbf{Final} & \textbf{Before} & \textbf{After} \\
\midrule
\small{MNIST} & 21 & 7 & 0.17s & 0.18s  \\
\small{MobileNet-V2-0.25} & 139 & 66 & 3.57s & 4s \\
\bottomrule
\end{tabular}
\end{table}
\section{Related Work} 
\label{sec:related}

Provenance has been studied across databases, workflow systems, machine learning platforms, and programming languages. Despite this breadth, existing approaches do not address a key challenge posed by modern AI compilers: tracking provenance through aggressive, non-injective graph rewrites without modifying compiler internals. We review prior work along three dimensions -- compiler transformations, ML workflow provenance, and invasive instrumentation -- and position our approach as a lightweight, observational alternative.

\paragraph{AI Compilers and Graph Rewriting}
Modern AI compilers -- including XLA~\cite{xla}, TVM~\cite{tvm}, Glow~\cite{glow}, and MLIR-based systems~\cite{mlir-llvm} -- optimize neural networks via repeated normalization, lowering, and hardware-specific rewrites. These transformations are semantics-preserving but structurally destructive: nodes may be fused, eliminated, duplicated, or reordered. Syntactic correspondence between pre- and post-pass graphs is often lost.

While these systems provide debugging utilities such as pass tracing and IR dumps, they do not offer a principled notion of provenance for tensors or operators across compilations. In particular, they lack a semantic account of how observable computation evolves under transformation.

Our work complements these systems by providing an external semantic model for provenance that does not rely on syntactic node correspondence. By reasoning about compilation in terms of observable computational actions rather than IR structure, our approach remains robust to aggressive rewrites while preserving the compiler’s existing transformation pipeline.

\paragraph{Provenance in Machine Learning Workflows}
A large body of work focuses on provenance for reproducibility, accountability, and governance in machine learning workflows. Systems such as those of Schlegel and Sattler~\cite{DBLP:conf/deem/SchlegelS23}, Ormenisan et al.~\cite{ormenisan2020implicit}, and yProv4ML~\cite{DBLP:journals/corr/abs-2507-01075} track datasets, training configurations, model artifacts, and evaluation results. Provenance is typically defined over experiment runs and artifacts, enabling replay, auditing, and comparison.

Similarly, workflow engines such as Apache Beam~\cite{beam}, Airflow~\cite{airflow}, and Kubeflow Pipelines~\cite{kubeflow} expose lineage via execution DAGs, logs, and versioned outputs. These systems emphasize operational reliability and end-to-end traceability across heterogeneous pipelines.

However, such approaches operate at a coarse granularity and treat models as opaque artifacts. They do not reason about the internal structure of computation graphs or the effects of compiler optimizations. Provenance is retrospective and metadata-oriented, rather than semantic and transformation-aware. In contrast, our work targets provenance within the compiler, focusing on how tensors and operators evolve across optimization passes rather than across experimental runs.

\paragraph{Invasive Provenance in ML Frameworks}
Some ML frameworks incorporate provenance-like mechanisms directly into their programming model. For example, TensorFlow extensions such as TFRecordDataset~\cite{tfrecord} encode lineage and metadata into data abstractions themselves. While effective for tracing data flow, these approaches are inherently invasive: they require changes to APIs, programming models, or intermediate representations. As a result, provenance becomes entangled with execution semantics and imposes significant engineering overhead.

Our approach deliberately avoids this trade-off. By treating provenance as an observation of transformations rather than an intrinsic property embedded in IR nodes, we achieve provenance stability without modifying generated code or intermediate representations at each pass.

\section{Conclusion}
\label{sec:conclusion}

This paper presents a coalgebraic framework for provenance-aware AI compilation. Our contributions are: (1) the formulation of \emph{observational correctness}, which preserves operator derivation histories through compiler transformations; (2) a generative model where provenance emerges as an observable invariant from rewrite traces; and (3) a practical implementation in the {\quartz} compiler that maintains full transformation histories without performance overhead.

This work connects compiler verification with critical needs in the ML community. Our approach enables new techniques in explainable AI (XAI), offering clear lineage from high-level operations to optimized code. This supports debugging, optimization auditing, and trustworthy ML toolchains -- making the \emph{why} behind compiled models as recoverable as their output.

\bibliographystyle{ACM-Reference-Format}
\bibliography{ref.bib}
\end{document}